\documentclass[12pt,dvips]{article}

\usepackage{epsfig,times,amssymb,amsmath,multirow,rotating}
\usepackage{picinpar}
\usepackage{wrapfig}
\usepackage{floatflt}
\makeatletter
\renewcommand{\section}{\@startsection{section}{1}{0in}
	{0.4\baselineskip}{0.1\baselineskip}{\Large\bf}}
\renewcommand{\subsection}{\@startsection{subsection}{2}{0in}
	{0.25\baselineskip}{-\baselineskip}{\large\bf}}
\renewcommand{\subsubsection}{\@startsection{subsubsection}{3}{0in}
	{0.1\baselineskip}{-\baselineskip}{\normalsize\bf}}
\makeatother
\newcommand{\bce}{\begin{center}}
\newcommand{\ece}{\end{center}}

\newcommand{\bal}{\begin{eqnarray*}}
\newcommand{\eal}{\end{eqnarray*}}
\newcommand{\bmat}{\begin{math}}
\newcommand{\emat}{\end{math}}
\newcommand{\bec}{\begin{displaymath}}
\newcommand{\eec}{\end{displaymath}}
\newcommand{\becnum}{\begin{equation}}
\newcommand{\eecnum}{\end{equation}}

\begin{document}
\begin{center}
{\Large \bf Fluxes of Atmospheric Leptons at 600 GeV - 60 TeV}

{\large
{\bf Dmitry Chirkin}\\
{\it chirkin@physics.berkeley.edu\\
University of California at Berkeley, USA}}

\end{center}

\begin{abstract}
\label{MMC}
The atmospheric muon flux is generated with CORSIKA 6.030 (using QGSJET high-energy interaction model) and fitted with the formula from Gaisser (1990). Spectra and mass distribution of primaries were taken according to the poli-gonato model of H\"orandel (2003). A convenient parameterization of the $\cos(\theta)\rightarrow\cos(\theta^*)$ correction, originally tabulated in Volkova (1969), is presented. This correction, together with muon energy losses and decay, is shown to significantly improve the fit to the simulated data.
\end{abstract}

\section{Introduction}
\label{intro}

The zenith-angle dependent energy distribution of muons, generated in showers initiated by high-energy cosmic rays in the atmosphere, can be conveniently parametrized \cite{dcors_8} as
\begin{equation}
\label{gaisser}
\frac{dN}{dE}={0.14 \over \mbox{ cm}^{2}\mbox{ sr}\mbox{ s}\mbox{ GeV}} \cdot \mbox{A} \cdot \left({E_{\mu} \over \mbox{GeV}}\right)^{-\gamma} \cdot \left(\frac{1}{1+\frac{1.1 E_{\mu}\cos (\theta) }{\mathrm{115\ GeV}}}+\frac{0.054}{1+\frac{1.1 E_{\mu}\cos (\theta) }{\mathrm{850\ GeV}}}\right)
\end{equation}
This formula is often used in the data analyses of the underground muon and neutrino detectors, e.g., for the vertical depth-intensity relation calculation or the estimates of the charm component contribution to the muon flux. The relative normalization $A$ and the spectral index $\gamma$ were varied in this work to fit the results of CORSIKA \cite{corsika} simulations. With the assumption of scaling in the high-energy hadron-nucleus interactions \cite{volkova79} used in derivation of this formula, the muon flux spectral index $\gamma$ should be the same as that for the primaries. However, in the parameterization of the cosmic ray energy spectrum and mass distribution used here \cite{horandel}, different primaries have different spectral indices. Therefore the expected value of the muon flux spectral index $\gamma$ is not immediately obvious, but should be close to some ``weighted average'' of spectral indices of the individual cosmic ray components. In \cite{kbs01} values of $\pi$ and $K$ critical energies \cite{volkovaZ} at $\theta=0$ and the ratio of $K$ to $\pi$-generated muons were also varied. In this work varying these parameters did not improve the fits, so they have been fixed at 115 GeV, 850 GeV, and 0.054 as indicated in the formula above (as in \cite{dcors_8}).

According to \cite{horandel}, the cosmic ray energy spectrum can be described by a sum of components with Z=1-92, each given by
$$d\Phi_Z/dE=\Phi_Z^0 E^{-\gamma_Z}\left[1+\left({E\over E_Z} \right)^{\epsilon_c} \right]^{-\Delta\gamma\over \epsilon_c}. $$
The best description of cosmic ray data within this model is achieved using rigidity-dependent cutoff energies $\quad E_Z=E_p\cdot Z$. Values of $E_p$, $\Delta\gamma$, and $\epsilon_c$, as well as a table of values of $\Phi_Z^0$ and $\gamma_Z$ used in this work are given in \cite{horandel}. Since CORSIKA can only treat primaries with Z=1-26, our cosmic ray spectrum implementation is only valid up to $\approx$ 50 PeV. As shown in \cite{horandel}, primaries with Z=28-92 as well as an additional ad-hoc component are important at energies above $\approx$ 100 PeV.

To determine the values of $A$ and $\gamma$ corresponding to this model of cosmic rays \cite{horandel}, a simulation based on CORSIKA version 6.030 was used. The high-energy interaction model QGSJET was shown to be the best to describe the muon fluxes observed by AMANDA \cite{mythesis} and several other experiments, and is also the fastest of the six high-energy interaction models available in CORSIKA. For these reasons it was used as the default model in this work. A detailed comparison with the results of other models is presented in \cite{mythesis}. A sample of $10^{10}$ showers with energies above 600 GeV was generated, which took approximately 200 GHz-CPU-days.

The muon flux formula given above (Equation \ref{gaisser}) was derived assuming a flat atmosphere, and is therefore only valid up to zenith angles of less than $70^{\circ}$. In order to describe the full range of zenith angles, $0^{\circ}-90^{\circ}$, the $\cos(\theta)$ dependence of the $\pi$ and $K$ critical energies valid at zenith angles below $70^{\circ}$ can be replaced with a $\cos(\theta^*)$ dependence as in \cite{volkovaZ}. In this work a convenient parameterization of this substitution is given.

\section{The $\cos(\theta)\rightarrow\cos(\theta^*)$ correction}
\label{costh}

\begin{figure}[!h]\begin{center}
\begin{tabular}{ccc}
\mbox{\epsfig{file=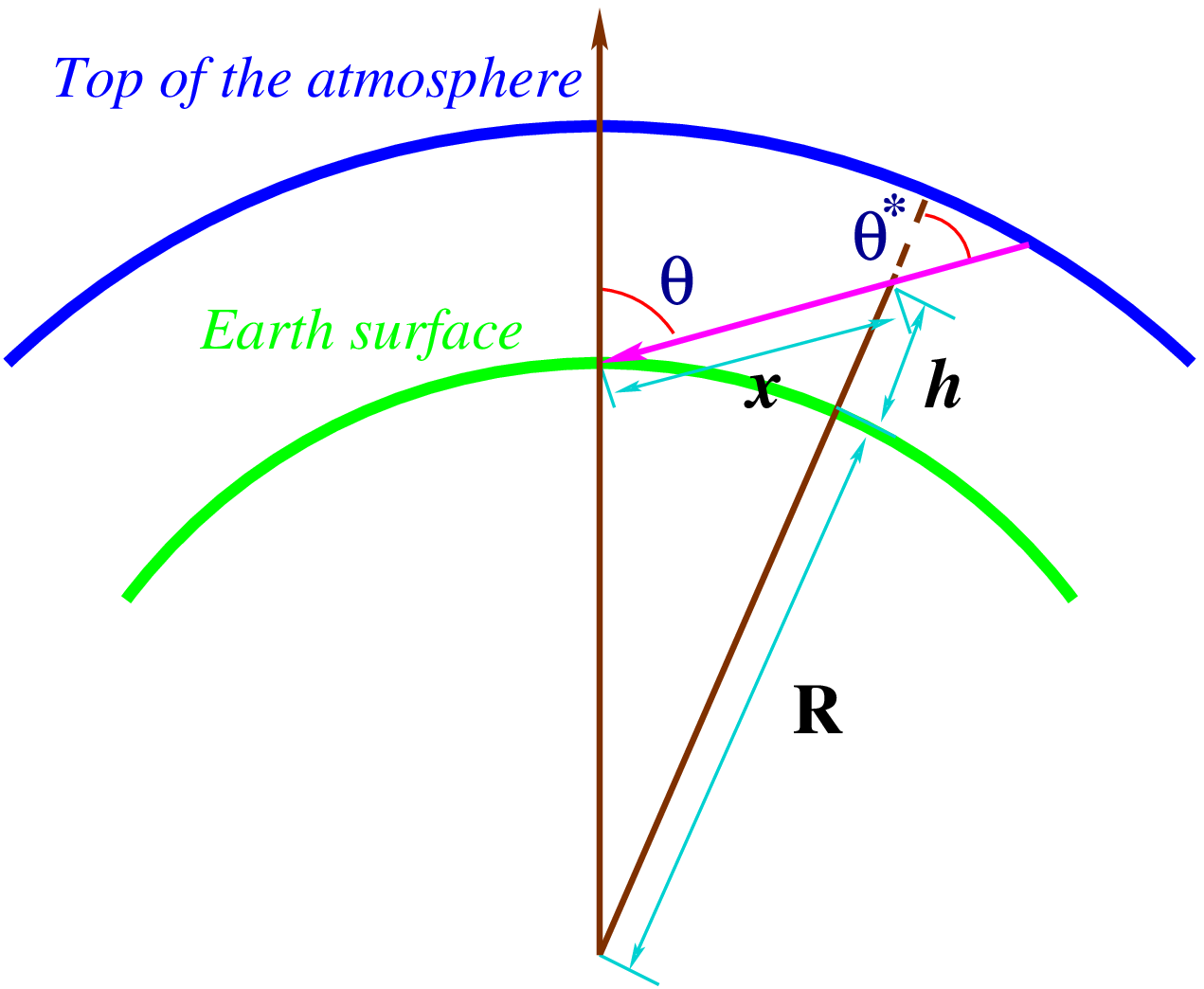,width=.45\textwidth}} & \ & \mbox{\epsfig{file=pics/zl.epsi,width=.45\textwidth}} \\
\parbox{.45\textwidth}{\caption[Model of the atmosphere. ]{\label{fig1}Model of the atmosphere. }} & \ & \parbox{.45\textwidth}{\caption[Muon track length. ]{\label{fig2}Muon track length. }} \\
\end{tabular}
\end{center}\end{figure}

The muon flux formula as given in the previous section (Equation \ref{gaisser}) cannot be used for $\theta>70^{\circ}$. In \cite{volkovaZ} the critical energies of $\pi$ and $K$, here given simply as $E_{cr}^{\pi}(\theta)=\mathrm{115\ GeV}/\cos(\theta)$ and $E_{cr}^K(\theta)=\mathrm{850\ GeV}/\cos(\theta)$, were calculated taking into account the curvature of the atmosphere, thus extending the range of applicability to all zenith angles. The definition of the critical energy of $\pi$ or $K$, i.e., the energy at which the $\pi$ or $K$ decay length is equal to the interaction length, is somewhat loose, and depends on the location of the first interaction of the cosmic ray primary. Its mean is estimated as $X_m=$85 g/cm$^2$ in \cite{volkovaZ}, and the average altitude of muon production as $h_0=$17 km (for the direction $\theta=0$) in \cite{LVD}. In \cite{volkova79} $X_m=(1+0.04\cdot \log_{10}(E/\rm{1\ TeV}))\cdot 85$ g/cm$^2$, i.e., $X_m$ has only a very weak energy dependence. In this work, values of $X_m=X_0 \log 2=$79.6 g/cm$^2$ and $h_0=$19.28 km are obtained by comparison with CORSIKA-simulated muon flux above 600 GeV for the standard US atmosphere parameters (Figure \ref{fig1}), as shown in Figure \ref{fig2}.

\begin{wrapfigure}{r}{5.5cm}
\epsfig{file=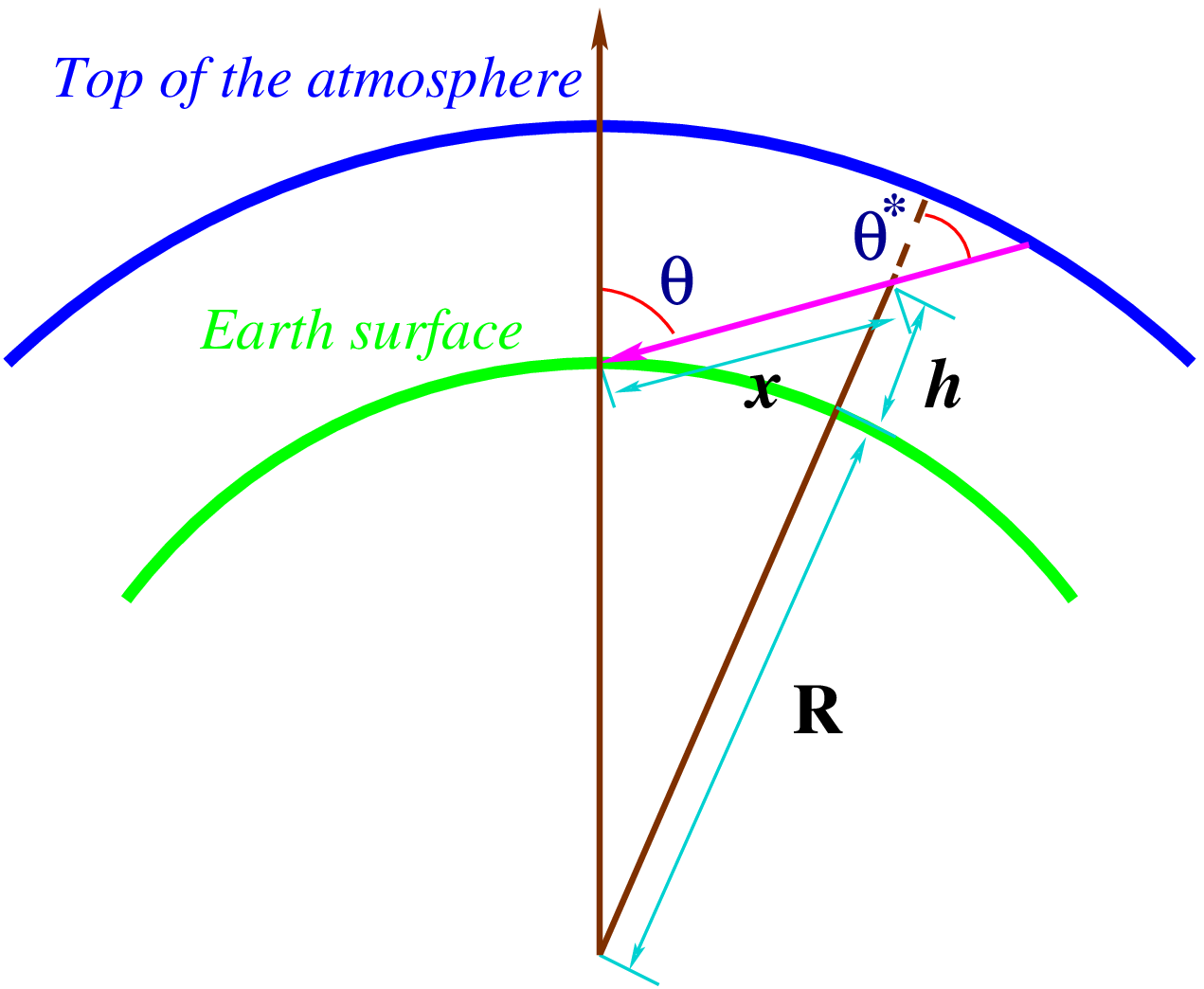,width=0.3\textwidth}
\end{wrapfigure}
Assuming that the interaction length of $\pi$ and $K$ in air
$$X_{int}^{\pi,K}= \int\limits_{x-\delta x_{\pi,K}}^x \rho dx \approx \rho(h(x)) \cdot \delta x_{\pi,K},$$
is approximately constant with energy, and the decay length is proportional to energy, $l_{\pi,k}= {\tau_{\pi,K} E_{\pi,K} \over m_{\pi,K} c}$, the critical energy (from $l_{\pi,k}=\delta x_{\pi,K}$) is $$E^{\pi,K}_{cr}={X_{int}^{\pi,K} m_{\pi,K}c \over \rho(h(x)) \tau_{\pi,K}} \propto {1\over \rho(h(x))}.$$
Therefore, $\cos(\theta^*)=E^{\pi,K}_{cr}(0)/E^{\pi,K}_{cr}(\theta)=\rho(h(x(\theta)))/\rho(h(x(0)))$.

At angles $\theta<70^{\circ}$ the atmosphere can be approximated as flat and $x=h/ \cos(\theta)$. The atmosphere above $\approx 200$ g/cm$^2$ can be considered isothermal \cite{dcors_8}, therefore the density $\rho(h)\propto \exp(-h/h_0)$ is proportional to the mass overburden above $h$, $h_0\rho(h)$. The first interaction of a cosmic ray primary on average occurs at $X \approx 85$ g/cm$^2$, where
\begin{equation}
\label{X}
X=\int\limits_x^{\infty}\rho(h(x))dx = \int\limits_h^{\infty}{\rho(h)\cdot dh \over \cos(\theta)} = {1\over \cos(\theta)} \int\limits_h^{\infty} \rho(h)\cdot dh = {h_0 \cdot \rho(h) \over \cos(\theta)},
\end{equation}
i.e. $\rho(h) \propto \cos(\theta)$. Therefore in this case $\cos(\theta^*)=\cos(\theta)$. At larger angles $\theta^*$ can be approximated as the zenith angle of the muon at the point of its production \cite{LVD}, as indicated in the figure above, since the integral in Equation \ref{X} is dominated by the contribution from the atmospheric layer with only a width $h_0 \approx 6.4$ km, which can be considered approximately flat.

The altitude of the first interaction, $h_i(\theta)$, can be determined from $X(\theta)=X(0)$, where
$$X(\theta)=\int\limits_{h_i(\theta)}^{\infty}\rho(h){dx \over dh}\ dh = \int\limits_{h_i(\theta)}^{\infty}{\rho (h)\ dh \over \sqrt{1- {\sin^2(\theta) \over (1+h/R)^2} }}, $$
where $R$ is the radius of the Earth. This equation was solved numerically for each $\theta$ and different $X=X(0)$. Then $\cos(\theta^*)=\rho(h_i(\theta))/\rho(h_i(0))$ as well as the muon track length from the production point $l(\theta)=\sqrt{R^2\cos^2(\theta)+2Rh_i+h_i^2}-R\cos(\theta)$ can be calculated. The quantities $h_i(\theta; X)$, $\cos(\theta^*; X)$, and $l(\theta; X)$ evaluated in this way are then averaged over muon production depth $X$ with weight $\exp(-X/X_0)/X_0$.

\begin{figure}[!b]\begin{center}
\begin{tabular}{ccc}
\mbox{\epsfig{file=pics/mpa1.epsi,width=.45\textwidth}} & \ & \mbox{\epsfig{file=pics/mpa2.epsi,width=.45\textwidth}} \\
\parbox{.45\textwidth}{\caption[Average muon production altitude. ]{\label{fig5}Average muon production altitude. }} & \ & \parbox{.45\textwidth}{\caption[$\cos(\theta^*)$ parameterization. ]{\label{fig6}$\cos(\theta^*)$ parameterization. }} \\
\end{tabular}
\end{center}\end{figure}

\begin{figure}[!h]\begin{center}
\begin{tabular}{ccc}
\mbox{\epsfig{file=pics/mpa4.epsi,width=.45\textwidth}} & \ & \mbox{\epsfig{file=pics/mpa3.epsi,width=.45\textwidth}} \\
\parbox{.45\textwidth}{\caption[Average muon track length. ]{\label{fig7}Average muon track length. }} & \ & \parbox{.45\textwidth}{\caption[Mass overburden of the atmosphere. ]{\label{fig8}Mass overburden of the atmosphere. }} \\
\end{tabular}
\end{center}\end{figure}

These averaged quantities can be fitted with
$$h_i(\theta)=p_5{1+p_3 \sin^{p_4}(\theta)\over \left[1- {\sin^2(\theta) \over (1+p_1)^2}  \right]^{p_2}}{\rm\ km},$$
$$\cos(\theta^*)=\sqrt{x^2+p_1^2+p_2x^{p_3}+p_4x^{p_5} \over 1+p_1^2+p_2+p_4},$$
$$\mbox{and} \quad l(\theta)={10^3{\rm\ km} \over p_1+p_2x^{p_3}+p_4(1-x^2)^{p_5}},$$
where $x=\cos(\theta)$. Additionally, the total atmospheric mass overburden as seen under zenith angle $\theta$ was parametrized as
$$X_{tot}(\theta)={1{\rm\ mwe} \over p_1+p_2x^{p_3}+p_4(1-x^2)^{p_5}}.$$

All fits were first performed assuming the last two parameters are zero, and then another time allowing all five parameters to vary, as indicated in Figures \ref{fig5}-\ref{fig8}. The five-parameter fits are accurate to within better than 0.40\% and their parameters are summarized in Table \ref{tab1}.

\begin{table}[!h]\begin{center}
\begin{tabular}{|ccccccc|}
\hline
fit & $p_1$ & $p_2$ & $p_3$ & $p_4$ & $p_5$ & av. deviation \\
$h_i(\theta)$ & 0.00851164 & 0.124534 & 0.059761 & 2.32876 & 19.279 & 0.04\% \\
$\cos(\theta^*)$ & 0.102573 & -0.068287 & 0.958633 & 0.0407253 & 0.817285 & 0.20\% \\
$l(\theta)$ & 1.3144 & 50.2813 & 1.33545 & 0.252313 & 41.0344 & 0.40\% \\
$X_{tot}(\theta)$ & -0.017326 & 0.114236 & 1.15043 & 0.0200854 & 1.16714 & 0.20\% \\
\hline
\end{tabular}
{\caption[Parameters of the fits to the quantities $h_i(\theta)$, $\cos(\theta^*)$, $l(\theta)$, and $X_{tot}(\theta)$ of Section \ref{costh}. ]{\label{tab1}Parameters of the fits to the quantities $h_i(\theta)$, $\cos(\theta^*)$, $l(\theta)$, and $X_{tot}(\theta)$ of Section \ref{costh}. }}
\end{center}\end{table}

The quantity $\cos(\theta^*)$ was also calculated as the zenith angle of the muon direction at the point of its production. As shown in Figure \ref{fig3}, the result is almost identical to the density ratio consideration. Also shown for comparison are the parameterizations of \cite{kbs01}.

\begin{figure}[!h]\begin{center}
\begin{tabular}{ccc}
\mbox{\epsfig{file=pics/costh.epsi,width=.45\textwidth}} & \ & \mbox{\epsfig{file=pics/loss.epsi,width=.45\textwidth}} \\
\parbox{.45\textwidth}{\caption[$\cos(\theta^*)$ parameterization. ]{\label{fig3}$\cos(\theta^*)$ parameterization. }} & \ & \parbox{.45\textwidth}{\caption[Muon energy loss parameters. ]{\label{fig4}Muon energy loss parameters. }} \\
\end{tabular}
\end{center}\end{figure}

\section{Muon energy losses and decay}
The muon integral flux at the muon production point $i$ is $$f_i(E_i)=\int\limits_{E_i}^{\infty}{dN_i \over dE}\ dE,$$ where $dN_i/dE$ is described by Equation \ref{gaisser} with the $\cos(\theta)$ correction of Section \ref{costh}. The muon integral flux at the surface, $f_f(E_f)$, can be calculated from
$$f_f(E_f)=\int\limits_{E_f}^{\infty} df_f(E)=\int\limits_{E_f}^{\infty} \langle df_i(E_i^{\xi}(E)) \rangle =\langle f_i(E_i^{\xi}(E_f)) \rangle,$$
where $E_i^{\xi}(E_f)$ is determined by a stochastic process of muon propagation and depends on a random variable $\xi$. To determine the function $f_f(E_f)$ one must find the average of $f_i(E_i^{\xi}(E_f))$ with respect to all possible muon initial energies $E_i^{\xi}$ which result in a muon final energy of $E_f$.

On average, final energy of $E_f$ corresponds to some $\langle E_i^{\xi} \rangle=E_f+\Delta E$. Using the usual linear approximation of the muon energy losses $dE/dX=a+bE$,
\begin{equation}
\label{dedx}
\langle E_i^{\xi}\rangle = ((a+bE_f)\exp(bX)-a)/b,
\end{equation}
where $X=X_{tot}(\theta)-X_0$ is the average mass overburden the muon crosses before reaching the surface. Assuming that the muon energy losses are small, the deviation of $E_i^{\xi}$ from its average $\langle E_i^{\xi} \rangle$ can also be estimated. Using an approximation $\langle (E_i^{\xi}-\langle E_i^{\xi}\rangle )^2\rangle =(a^*+b^*E)^2dX$ on a small path $dX$, and assuming that such deviations can be added on a path $X$, one derives 
\begin{equation}
\label{d2edx2}
\langle (E_i^{\xi}-\langle E_i^{\xi}\rangle)^2\rangle\approx\int\limits_{X_f}^{X_i}(a^*+b^*E)^2dX=\int\limits_{E_f}^{E_i}{(a^*+b^*E)^2\over a+bE}dE.
\end{equation}
Formulae \ref{dedx} and \ref{d2edx2} were used to fit a Monte Carlo simulation of muons propagating with energies $\sim$ 600 GeV $-$ 60 TeV through 10 and 360 mwe of air performed with MMC \cite{mmc}. Results of the fits are shown in Figure \ref{fig4}. Dashed blue lines show the result of muon propagation through 10 mwe, and solid green lines through 360 mwe, both rescaled to 1 mwe. The region of the fit is shown with vertical lines and the fits themselves with solid light blue lines. A single set of parameters $a^*$ and $b^*$ seems to describe well the $\langle (E_i^{\xi}-\langle E_i^{\xi}\rangle)^2\rangle$ deviations for both 10 and 360 mwe thus justifying the assumption of their additivity.

It is now possible to derive the expression for $f_f(E_f)$:
$$f_f(E_f)=\langle f_i(E_i^{\xi}(E_f))\rangle=\langle f_i(\langle E_i^{\xi}\rangle+(E_i^{\xi}-\langle E_i^{\xi}\rangle ))\rangle$$
$$\approx f_i(\langle E_i^{\xi}\rangle)+{1 \over 2}{d^2f\over dE^2}(\langle E_i^{\xi}\rangle) \langle(E_i^{\xi}-\langle E_i^{\xi}\rangle )^2\rangle$$
$$\approx f_i\left(\langle E_i^{\xi}\rangle+{1 \over 2}{d^2f/dE^2\over df/dE}(\langle E_i^{\xi}\rangle ) \langle (E_i^{\xi}-\langle E_i^{\xi}\rangle)^2\rangle\right)=f_i(E_I(E_f)).$$
Therefore, to account for the muon decay it is sufficient to replace $E_i$ in the formula from Section \ref{intro} with $E_I(E_f)$. Additionally, $dE_i\rightarrow (dE_I/dE_f)dE_f$. All quantities of this section (including the derivative $dE_I/dE_f$) can easily be evaluated and are implemented in the function fitted to the muon spectrum.

Muon decay is evaluated as a correction factor $\exp(-l\mu c/\tau E_I)$.

\section{Fits to the muon energy spectrum at the surface}
\label{fits}
The fit to Formula \ref{gaisser} is shown in Figure \ref{fig9}, and to the corrected formula in Figure \ref{fig10}. Both figures show parameters of the fit $A$ and $\gamma$ as well as the $\chi^2$ of the fit in the region of $\cos(\theta)=0.3-1$ and in the full range $\cos(\theta)=0-1$. Figure \ref{fig14} shows the behavior of the parameters and $\chi^2$ of the fit as a function of the lower boundary of the fit, the upper boundary always at $\cos(\theta)=1$. Equation \ref{gaisser} can only be used for $\cos(\theta)>0.3$, as mentioned before, and gives the same result in this region as the $\cos(\theta)$-only corrected formula. Both approaches fail to produce values of the fit parameters independent of the region of the fit. Including muon energy losses substantially improves the fit. Additionally correcting for muon decay gives the best description of the muon spectra. The value of $\chi^2$ drops to 1.2 at lower boundary of $\cos(\theta_{l})=0.03$ and is 1.12 at $\cos(\theta_{l})=0.3$. The values of $A$ and $\gamma$ vary very little as the region of the fit changes, and are $0.7015$ and $2.715$ at $\cos(\theta_{l})=0.3$.
\vspace{-0.3cm}
\begin{figure}[!b]\begin{center}
\begin{tabular}{ccc}
\mbox{\epsfig{file=pics/a12std000.3.epsi,width=.45\textwidth}} & \ & \mbox{\epsfig{file=pics/a12std300.3.epsi,width=.45\textwidth}} \\
\parbox{.45\textwidth}{\caption[Fit with Formula \ref{gaisser} to the CORSIKA-simulated muon energy spectrum. ]{\label{fig9}Fit with Formula \ref{gaisser} to the CORSIKA-simulated muon energy spectrum. }} & \ & \parbox{.45\textwidth}{\caption[Fit with the corrected Formula \ref{gaisser}. ]{\label{fig10}Fit with the corrected Formula \ref{gaisser}. }} \\
\end{tabular}
\end{center}\end{figure}

The resulting muon differential flux at 10 TeV is plotted in Figure \ref{fig13}, which can be compared with a figure containing results of different approaches to calculation of $\cos(\theta^*)$ from \cite{LVD}. Resulting fits to the muon integral fluxes are shown in Figures \ref{fig11} and \ref{fig12}. CORSIKA-simulated muon energy spectra appear to have some unexpected structure at $\cos(\theta)<0.03$, which causes a high $\chi^2\approx50$ of the fit to the whole region of $\cos(\theta)=0-1$. This was noticed before \cite{dcors_update}, but was shown not to affect the simulation of the muon flux at the depth of AMANDA-II (1730 m) or deeper. The highest value of the zenith angle at the surface which can be seen at the depth of AMANDA-II is 88.7$^{\circ}$, i.e. $\cos(\theta)>0.023$, where the artifact is much less pronounced.

\begin{figure}[!h]\begin{center}
\begin{tabular}{ccc}
\mbox{\epsfig{file=pics/fit1.epsi,width=.45\textwidth}} & \ & \mbox{\epsfig{file=pics/b.epsi,width=.45\textwidth}} \\
\parbox{.45\textwidth}{\caption[Behavior of the parameters of the fit. ]{\label{fig14}Behavior of the parameters of the fit. }} & \ & \parbox{.45\textwidth}{\caption[Muon differential flux at 10 TeV. ]{\label{fig13}Muon differential flux at 10 TeV. }} \\
\end{tabular}
\end{center}\end{figure}

\begin{figure}[!h]\begin{center}
\begin{tabular}{ccc}
\mbox{\epsfig{file=pics/z1.epsi,width=.45\textwidth}} & \ & \mbox{\epsfig{file=pics/z2.epsi,width=.45\textwidth}} \\
\parbox{.45\textwidth}{\caption[Integrated muon flux at 600 GeV. ]{\label{fig11}Integrated muon flux at 600 GeV. }} & \ & \parbox{.45\textwidth}{\caption[Integrated muon flux at 10 TeV. ]{\label{fig12}Integrated muon flux at 10 TeV. }} \\
\end{tabular}
\end{center}\end{figure}

\section{Uncertainties in the fit model}

To investigate the effect of the uncertainties in the knowledge of the atmospheric profile, the standard US atmosphere used in the fits of Section \ref{costh} was replaced with the winter (July) atmosphere at the South Pole (see Figure \ref{fig1} for comparison with standard US atmosphere). The ground level for this South Pole atmosphere was chosen at 115.5 m below the see level to match the atmospheric pressure at ground with that of the Standard US atmosphere.  As seen from Figures \ref{fig5}-\ref{fig8}, only the average muon production altitude and the muon track length undergo substantial changes. As seen from Figure \ref{fig6}, the change in the average muon production does not affect the $\cos(\theta^*)$ calculation. Most of the muon decays occur for muons with the longest tracks, which come from near the horizon. There the deviation in the track length from that calculated for the standard US atmosphere is small, as seen from Figure \ref{fig7}. Therefore, quantities parameterized in Section \ref{costh} for the standard US atmosphere can be used for the atmosphere with quite different profile, in the extreme case the winter atmosphere at the South Pole.

Fits of the previous section were calculated to the CORSIKA-simulated muon data generated in the absence of the magnetic field. As seen from Figure \ref{fig18}, adding a 50 nT magnetic field changes the muon scattering profile. The scattering width is 2-4 times that without the magnetic field, but is still very small for muons with energies above 600 GeV.

Fit parameters for both a wrong atmospheric profile and a non-zero magnetic field are shown in Figure \ref{fig17}. As expected, the change from the calculation of Section \ref{fits} is negligible. However, the parameters do change somewhat when the fits are recalculated in the energy range 1-100 TeV.

\begin{figure}[!h]\begin{center}
\begin{tabular}{ccc}
\mbox{\epsfig{file=pics/fit2.epsi,width=.45\textwidth}} & \ & \mbox{\epsfig{file=pics/ang.epsi,width=.45\textwidth}} \\
\parbox{.45\textwidth}{\caption[Uncertainties in the fit parameters. ]{\label{fig17}Uncertainties in the fit parameters. }} & \ & \parbox{.45\textwidth}{\caption[Magnetic field influence. ]{\label{fig18}Magnetic field influence. }} \\
\end{tabular}
\end{center}\end{figure}

\section{Energy spectra of atmospheric muon and electron neutrinos}

Along with muons, CORSIKA generates muon and electron neutrinos, spectra of which were fitted with the formulae from \cite{dcors_9}. Muon neutrino spectra were fitted with
$$ \frac{dN}{dE}={2.85\cdot 10^{-2} \over \mbox{ cm}^{2}\mbox{ sr}\mbox{ s}\mbox{ GeV}} \cdot \mbox{A} \cdot \left({E_{\nu_{\mu}} \over \mbox{GeV}}\right)^{-\gamma} \cdot \left(\frac{1}{1+\frac{6 E_{\nu_{\mu}}\cos (\theta^*) }{\mathrm{121\ GeV}}}+\frac{0.213}{1+\frac{1.44 E_{\nu_{\mu}}\cos (\theta^*) }{\mathrm{897\ GeV}}}\right), $$
and electron neutrino spectra were fitted with
$$ \frac{dN}{dE}={2.4\cdot 10^{-3} \over \mbox{cm}^{2}\mbox{ sr}\mbox{ s}\mbox{ GeV}} \cdot \mbox{A} \cdot \left({E_{\nu_e} \over \mbox{GeV}}\right)^{-\gamma} \cdot \left(\frac{0.05}{1+\frac{1.5 E_{\nu_e}\cos (\theta^*) }{\mathrm{897\ GeV}}}+\frac{0.185}{1+\frac{1.5 E_{\nu_e}\cos (\theta^*) }{\mathrm{194\ GeV}}}+\frac{11.4E_{\nu_e}^{\zeta(\theta)}}{1+\frac{1.21 E_{\nu_e}\cos (\theta^*) }{\mathrm{121\ GeV}}}\right), $$
$${\rm where} \quad \zeta=a+b\log_{10}E,\quad
a=
\left\{
\begin{array}{l}
0.11-2.4\cos(\theta)\\
-0.46-0.54\cos(\theta)
\end{array},
\right.
b=
\left\{
\begin{array}{lc}
-0.22+0.69\cos(\theta) \quad & \cos(\theta)<0.3\\
-0.01+0.01\cos(\theta) \quad & \cos(\theta)\geqslant 0.3
\end{array}.
\right.
$$

The muon neutrino energy spectrum fit demonstrates the same stability with respect to changes in the lower $\cos(\theta_l)$ boundary of the fit as the muon energy spectrum fit. Electron neutrino fit parameters, however, change in the range $A=0.732-0.998$ and $\gamma=2.689-2.732$ as $\cos(\theta_l)$ is varied from 0 to 0.5. The fits shown in Figures \ref{fig15} and \ref{fig16} were produced with $\cos(\theta_l)=0.3$.

\begin{figure}[!h]\begin{center}
\begin{tabular}{ccc}
\mbox{\epsfig{file=pics/a12num100.3.epsi,width=.45\textwidth}} & \ & \mbox{\epsfig{file=pics/a12nue100.3.epsi,width=.45\textwidth}} \\
\parbox{.45\textwidth}{\caption[Muon neutrino energy spectrum fit. ]{\label{fig15}Muon neutrino energy spectrum fit. }} & \ & \parbox{.45\textwidth}{\caption[Electron neutrino energy spectrum fit. ]{\label{fig16}Electron neutrino energy spectrum fit. }} \\
\end{tabular}
\end{center}\end{figure}

\section{Results and Conclusions}
Muon and muon- and electron neutrino energy spectra produced with CORSIKA in the energy range 600 GeV $-$ 60 TeV were fitted with Formula \ref{gaisser} after a $\cos(\theta)$ correction as described in Section \ref{costh} (see Table \ref{tab2}). Muon spectra fits were additionally corrected for the muon energy losses and decay. These corrections were shown to be important in the selected energy range. Convenient parameterizations of $\cos(\theta)$ correction and muon energy losses were given.

\begin{table}[!h]\begin{center}
\begin{tabular}{|ccccccc|}
\hline
 & $\mu$ & $\nu_{\mu}$ & $\nu_e$ & $\mu^-/\mu^+$ & $\nu_{\mu}/\bar{\nu}_{\mu}$ & $\nu_e/\bar{\nu}_e$ \\
$A$ & 0.701 $\pm$ 0.006 & 0.646 $\pm$ 0.003 & 0.828 $\pm$ 0.155 & 0.340/0.367 & 0.352/0.310 & 0.465/0.472 \\
$\gamma$ & 2.715 $\pm$ 0.001 & 2.684 $\pm$ 0.001 & 2.710 $\pm$ 0.023 & 2.720/2.712 & 2.680/2.696 & 2.719/2.741 \\
$A$ & \multicolumn{3}{c}{with $\gamma$ as given above} & 0.330/0.374 & 0.362/0.285 & 0.437/0.382 \\
\hline
\end{tabular}
{\caption[Muon and muon- and electron neutrino energy spectrum parameters. ]{\label{tab2}Muon and muon- and electron neutrino energy spectrum parameters. }}
\end{center}\end{table}

\end{document}